\documentclass[preprintnumbers,amsmath,amssymb,floatfix,11pt,prd,onecolumn,superscriptaddress,nofootinbib]{revtex4}
\usepackage{latexsym}
\usepackage{epsfig}
\usepackage{amsmath}
\usepackage{epstopdf}
\usepackage{amssymb}
\usepackage{graphicx}
\usepackage{hyperref}
\usepackage{tabularx}
\usepackage{varioref}
\usepackage{placeins}
\newcolumntype{C}[1]{>{\centering\arraybackslash}p{#1}}

\begin{document}
\title{\bf Center of Mass Energy of the Collision for Two General Geodesic Particles Around a Kerr-Newman-Taub-NUT Black Hole}

\author{Ayesha Zakria}
\email{ayesha.zakria@sns.nust.edu.pk}
\affiliation{School of Natural Sciences, National University of
Sciences and Technology, Islamabad, Pakistan}

\author{Mubasher Jamil}
\email{mjamil@sns.nust.edu.pk}
\affiliation{School of Natural Sciences, National University of
Sciences and Technology, Islamabad, Pakistan}

\begin{abstract}
{\bf Abstract:} In this paper, we investigate the center of mass energy of the collision for two neutral particles with different rest masses falling
freely from rest at infinity in the background of a Kerr-Newman-Taub-NUT black hole. Further, we discuss the center of mass energy near the horizon(s) of an extremal and non-extremal Kerr-Newman-Taub-NUT black hole and show that an arbitrarily high center of mass energy is achievable under some restrictions. 
\end{abstract}
 \maketitle
 \newpage
\section{Introduction}
Collision of particles frequently take place in the accretion disks around black holes. The phenomenon of arbitrary high CME is a purly relativistic phenomenon and they serves as an excellent tool to study high energy astrophysics. Ba\~{n}ados, Silk and West (BSW) \cite{1} studied the collision for two
particles around a Kerr black hole and determined the center of mass energy (CME) in the equatorial plane. Subsequently, in \cite{2, 3}, the authors further elucidated the BSW mechanism. They pointed out that the arbitrarily high CME might not be
achievable in nature due to the astrophysical limitations i.e., the maximal spin and gravitational radiation.
Lake \cite{4,5}  demonstrated that the CME for two colliding particles is divergent at the inner horizon of a non-extremal Kerr black hole. Grib and Pavlov
\cite{6}-\cite{8} showed that very large values of the scattering energy of particles in the centre of mass
frame can be obtained for an extremal and non-extremal Kerr
black hole. The collision in the innermost stable circular orbit for a Kerr black hole
was discussed in \cite{9}. In \cite{10}, the author considered the collision for two neutral particles within the context of the near-horizon extremal Kerr black hole and demonstrated that the CME is finite for any admissible value of the particle parameters. In \cite{11}, the authors showed that the particle acceleration to arbitrary high energy is one of the universal properties of an extremal Kerr black hole not only in astrophysics but also in more general context. An explicit expression of the CME for two colliding general
geodesic massive and massless particles at any spacetime point around a Kerr black hole was obtained in \cite{12}. They found that, in the direct collision scenario, an arbitrarily high CME can
arise near the horizon of an extremal Kerr black hole not only at the equator but also on a belt
centered at the equator. This belt lies between latitudes $\pm a\cos(\sqrt{3}-1)\simeq \pm42.94^{\circ}$. In \cite{13}, the author argued the possibility of having infinite CME in the centre of mass frame of
colliding particles is a generic feature of a Kerr black hole.

In \cite{14}, the authors investigated the CME in the background of a Kerr-Newman black hole. They pointed out that the unlimited CME requires three conditions: (1) the collision takes place at the horizon of an extremal black hole, (2) one of the colliding particles has critical angular momentum, and (3) the spin parameter $a$ satisfies {\Large $\frac{1}{\sqrt{3}}$}$\leq a\leq 1$. In \cite{15}, the author studied the collision of two general geodesic particles around a Kerr-Newman black hole and
get the CME of the non-marginally and marginally bound critical particles. The collision for a freely falling neutral particle with a charged particle revolving in the circular orbit around a Schwarzschild black hole was considered in \cite{16}. In \cite{17}, the authors studied the collision for two particles with different rest masses moving in the equatorial plane of a Kerr-Taub-NUT black hole. They demonstrated that the CME depends on the spin parameter $a$ and NUT (Newman-Unti-Tamburino) charge $n$. Exact Lense-Thirring (LT) precession and causal geodesics in the inner-most stable circular orbit (ISCO) in a Kerr-Taub-NUT black hole was studied in \cite{18,19,20}. The CME of the collision for two uncharged particles falling freely from rest at infinity in the background of a charged, rotating and accelerating black hole was investigated in \cite{21}.

In \cite{22}, the authors discussed the collision for two particles in the background of a stringy black hole. They found that the CME is arbitrarily high under two conditions: (1) the spin parameter $a\neq0$, and (2) one of the colliding particles should have critical angular momentum. The collision for two particles in the background of a charged black string was discussed in \cite{23}. It was shown that the CME is arbitrarily high at the outer horizon if one of the colliding particles has critical charge. The particle acceleration mechanism five-dimensional compact black string, has been studied in \cite{24}. They found that the scattering energy of particles in the center of mass frame can take arbitrarily large values not only for an extremal black string but also for a non-extremal black string. The CME in the absence and presence of a magnetic field around a Schwarzschild-like black hole was investigated in \cite{25}. In \cite{26}, the authors discussed the CME for two colliding neutral particles at the horizon of a slowly rotating black hole in the Horava-Lifshitz theory of gravity and a topological Lifshitz black hole remains finite. The collision for test charged particles in the vicinity of the event horizon of a weakly magnetized static black hole with gravitomagnetic charge studied in \cite{27}. In \cite{28}, the author argued that the BSW effect exists for a non-rotating but charged black hole even for the simplest case of radial motion of
particles in a Reissner-Nordstr\"{o}m black hole. In \cite{29}, the author gave simple and general explanation to the effect of unbound acceleration of particles for Reissner-Nordstr\"{o}m and Kerr black holes. The CME of the collision for charged particles in a Bardeen black hole was studied in \cite{30}. In \cite{31}, the authors investigated the CME near the horizon of a non-extremal Plebanski-Demianski black hole without NUT parameter. The CME in the background of Ay\`{o}n-Beato-Garc\`{i}a-Bronnikov (ABGB), Einstein-Maxwell-dilaton-axion (EMDA) and  Ba\~{n}ados-Teitelboim-Zanelli (BTZ) black holes was investigated in \cite{32}.

A non-vacuum solution of the Einstein field equations is a Kerr-Newman-Taub-NUT (KNTN) black hole,
which besides the spin parameter $a$ and electric charge $Q$ carries the NUT charge $n$, the later one plays the role of a magnetic charge. We adopt the Hamilton-Jacobi approach to study the dynamics of a neutral particle in the background of a KNTN black hole. We do not restrict the dynamics and collision to the equatorial plane alone. Instead we choose arbitrary $\theta$ and fix $\theta=$ {\Large $\frac{\pi}{2}$} only as a special case. We discuss the detailed behavior of the CME for two neutral particles with different rest masses $m_{1}$ and $m_{2}$ falling freely from rest at infinity in the background of a KNTN black hole. We derive the CME when the collision occurs at some radial coordinate $r$ and angle $\theta$ close to the horizon. We show that the CME near the horizon(s) of an extremal and non-extremal KNTN black hole is arbitrarily high when the specific angular momentum of one of the colliding particles is equal to the critical angular momentum and non-vanishing spin parameter $a$.

The paper is organized as follows. In Sec. II, we will discuss the equations of motion for a neutral particle in the background of a KNTN black hole. In Sec. III, we will obtain the CME of the collision for two neutral particles and discuss the properties. In Sec. IV, we will give a brief conclusion. We use the system of units $c=G=1$ throughtout this paper.

\section{Equations of motion in the background of a Kerr-Newman-Taub-NUT black hole}
In this section, we will study the equations of motion for a neutral particle in the background of a KNTN black hole. Let us first give a brief review of a KNTN black hole. The KNTN black hole is a geometrically stationary and axisymmetric non-vacuum object, which is an important solution of the Einstein field equations. The KNTN black hole is determined by the following parameters i.e., the mass $M$, spin parameter $a$, NUT parameter $n$ and electric charge $Q$. The KNTN black hole can be described by the metric in the Boyer-Lindquist coordinates $(t, r, \theta, \phi)$ as in \cite{33,34,35}
\begin{eqnarray}\label{1}
ds^{2}&=&-\frac{1}{\Sigma}(\Delta-a^{2}\sin^{2}\theta)dt^{2}+\frac{2}{\Sigma}\big(\chi\Delta-a(\Sigma+a\chi)\sin^{2}\theta\big)dtd\phi+\frac{1}{\Sigma}\big((\Sigma+a\chi)^{2}\sin^{2}\theta-\chi^{2}\Delta\big)d\phi^{2}\notag\\&&
+\frac{\Sigma}{\Delta}dr^{2}+\Sigma d\theta^{2},
\end{eqnarray}
where $\Sigma$, $\Delta$ and $\chi$ are respectively defined by
\begin{eqnarray}\label{2}
\Sigma&=&r^{2}+(n+a\cos\theta)^{2}, \notag\\     \Delta &=&r^{2}-2Mr-n^{2}+a^{2}+Q^{2},\\ \chi &=&a\sin^{2}\theta-2n\cos\theta\notag.
\end{eqnarray}
The KNTN metric contains the following metrics as special cases: Kerr-Taub-NUT $(Q=0)$, Taub-NUT $(a=Q=0)$, Kerr-Newman $(n=0)$, Reissner-Nordstr\"{o}m $(a=n=0)$, Kerr $(n=Q=0)$ and Schwarzschild $(a=n=Q=0)$.

The metric (\ref{1}) becomes singular if $\Sigma=0$ or $\Delta=0$, whereas $\Sigma=0$ is the curvature singularity and $\Delta=0$ is the coordinate singularity\footnote{The curvature invariants are given in Appendix.}. Here, $\Sigma=0$ implies $r=0$ and $\cos\theta=$ {\Large $-\frac{n}{a}$}. The horizon(s) of the KNTN black hole occur at $ r_{\pm}=M\pm\sqrt{M^{2}+n^{2}-a^{2}-Q^{2}}$,
where $r_{+}$ and $r_{-}$ define the outer and inner horizons, respectively, which are roots of the equation $\Delta=0$. The existence of the horizons require $n^{2}\geq a^{2}+Q^{2}-M^{2}$, where ``$=$" and ``$>$" correspond to the extremal and non-extremal KNTN black holes, respectively.

Now, let us discuss the equations of motion for a neutral particle of mass $m$ in the background of a KNTN black hole. The motion of the particle can be determined by the Lagrangian
\begin{equation}\label{3}
 \mathcal{L}=\frac{1}{2}g_{\mu\nu}\dot{x}^{\mu}\dot{x}^{\nu},
\end{equation}
where the overdot denotes differentiation with respect to an affine parameter $\lambda$ related to the proper time $\tau$ by $\tau=m\lambda$.
The normalization condition is {\Large $\frac{1}{m^2}$} $g_{\mu\nu}\dot{x}^{\mu}\dot{x}^{\nu}=\kappa$, where $\kappa=-1$ for timelike geodesics, $\kappa=0$ for null geodesics and $\kappa=1$ for spacelike geodesics. For the massive particle, we have $\kappa=-1$. The 4-momentum of the particle is
\begin{equation}\label{4}
  P_{\mu}=\frac{\partial \mathcal{L}}{\partial \dot{x}^{\mu}}=g_{\mu\nu}\dot{x}^{\nu},
\end{equation}
which is related to the 4-velocity by
 \begin{equation}\label{5}
 u_{\mu}=\frac{P_{\mu}}{m},
\end{equation}
where $u^{\nu}=$ {\Large $\frac{dx^{\nu}}{d\tau}$}, $\tau$ is the proper time for timelike geodesics. Using Eq. (\ref{4}), we can express $\dot{x}^{\mu}$ in terms of the 4-momentum as $\dot{x}^{\mu}=g^{\mu\nu}P_{\nu}$. The Hamiltonian is given by
\begin{equation}\label{6}
  \mathcal{H}=P_{\mu}\dot{x}^{\mu}-\mathcal{L}=\frac{1}{2}g^{\mu\nu}{P}_{\mu}{P}_{\nu},
\end{equation}
which satisfies the Hamilton equations
\begin{equation}\label{7}
\dot{x}^{\mu}=\frac{\partial \mathcal{H}}{\partial P_{\mu}},~\dot{P}_{\mu}=-\frac{\partial \mathcal{H}}{\partial x^{\mu}}.
\end{equation}
Moreover, the Hamilton-Jacobi equation is given by
\begin{equation}\label{8}
  \mathcal{H}=-\frac{\partial S}{\partial \lambda}=\frac{1}{2}g^{\mu\nu}\frac{\partial S}{\partial x^{\mu}}\frac{\partial S}{\partial x^{\nu}},
\end{equation}
where $S$ is the Jacobi action and
\begin{equation}\label{9}
  \frac{\partial S}{\partial x^{\mu}}=P_{\mu}.
\end{equation}
The Hamilton-Jacobi equation allows separation of variables in the form
\begin{equation}\label{10}
S(t,r,\theta,\phi) =\frac{1}{2}m^2\lambda-\mathcal{E}t+\mathcal{L}\phi+S_{r}(r)+S_{\theta}(\theta),
\end{equation}
where $\mathcal{E}$ and $\mathcal{L}$ are respectively the energy and angular momentum of the particle, $S_{r}$ and $S_{\theta}$ are arbitrary functions of $r$ and $\theta$, respectively. Here, {\Large $\frac{1}{m}$}{\Large $\frac{\partial S}{\partial t}$} $=-E$ and
{\Large $\frac{1}{m}$}{\Large $\frac{\partial S}{\partial \phi}$} $=L$, where $E$ and $L$ are the specific energy and specific angular momentum of the particle defined by $E=${\Large $\frac{\mathcal{E}}{m}$} and $L=${\Large $\frac{\mathcal{L}}{m}$}. Using these relations and Eq. (\ref{9}), we get
\begin{eqnarray}
 E&=&-\frac{P_{t}}{m}=\frac{\Delta-a^{2}\sin^{2}\theta}{\Sigma}u^{t}-\frac{\chi\Delta-a(\Sigma+a\chi)\sin^{2}\theta}{\Sigma}u^{\phi},\label{11}\\
 L&=&\frac{P_{\phi}}{m}=\frac{\chi\Delta-a(\Sigma+a\chi)\sin^{2}\theta}{\Sigma}u^{t}+\frac{(\Sigma+a\chi)^{2}\sin^{2}\theta-\chi^{2}\Delta}{\Sigma}u^{\phi}.\label{12}
\end{eqnarray}
Solving Eqs. (\ref{11}) and (\ref{12}), we obtain
\begin{eqnarray}
u^{t}&=&\frac{\chi(L-\chi E)}{\Sigma\sin^{2}\theta}+\frac{(\Sigma+a\chi)\big[E(\Sigma+a\chi)-aL\big]}{\Delta\Sigma},\label{13}\\
u^{\phi}&=&\frac{L-\chi E}{\Sigma\sin^{2}\theta}+\frac{a\big[(\Sigma+a\chi)E-aL\big]}{\Delta\Sigma}.\label{14}
\end{eqnarray}
By Eqs. (\ref{8}) and (\ref{10}), we obtain
\begin{eqnarray}\label{15}
\frac{1}{m^2}\bigg(\frac{\partial S_{\theta}}{\partial \theta}\bigg)^{2}+\cos^{2}\theta\bigg(\big(1-E^{2}\big)a^{2}+\frac{L^{2}}{\sin^{2}\theta}\bigg)+2an\cos\theta\big(1-2E^{2}\big)
+\frac{4n\cos\theta E}{\sin^{2}\theta}\Big(n\cos\theta E+L\Big)
\notag\\=-\frac{\Delta}{m^2}\bigg(\frac{\partial S_{r}}{\partial r}\bigg)^{2}-r^{2}-n^{2}-\big(L-aE\big)^{2}+\frac{1}{\Delta}\Big(\big(r^{2}+n^{2}+a^{2}\big) E-aL\Big)^{2}.~~~~~~~~~~~~~~
\end{eqnarray}
The left-hand side of Eq. (\ref{15}) does not depend on $r$ while the right-hand side does not depend on $\theta$, hence each side must be a constant. This constant is termed as the Carter constant denoted by $K$ and is a conserved quantity. Therefore
\begin{equation}\label{16}
\frac{1}{m^2}\bigg(\frac{\partial S_{\theta}}{\partial \theta}\bigg)^{2}+\cos^{2}\theta\bigg(\big(1-E^{2}\big)a^{2}+\frac{L^{2}}{\sin^{2}\theta}\bigg)+2an\cos\theta\big(1-2E^{2}\big)
+\frac{4n\cos\theta E}{\sin^{2}\theta}\Big(n\cos\theta E+L\Big)=K,
\end{equation}
\begin{equation}\label{17}
\frac{\Delta}{m^2}\bigg(\frac{\partial S_{r}}{\partial r}\bigg)^{2}+r^{2}+n^{2}+\big(L-aE\big)^{2}-\frac{1}{\Delta}\Big(\big(r^{2}+n^{2}+a^{2}\big) E-aL\Big)^{2}=-K.
\end{equation}
Using the relations $u_{r}=$ {\Large $\frac{1}{m}$}{\Large $\frac{\partial S_{r}}{\partial r}$} and $u_{\theta}=$ {\Large $\frac{1}{m}$}{\Large $\frac{\partial S_{\theta}}{\partial \theta}$}, the remaining 4-velocity components are
\begin{eqnarray}
\Sigma u^{\theta}&=&\pm\sqrt{\Theta},\label{18}\\
\Sigma u^{r}&=&\pm\sqrt{R},\label{19}
\end{eqnarray}
with
\begin{eqnarray}
\Theta=\Theta(\theta)&=&K-\cos^{2}\theta\bigg((1-E^{2})a^{2}+\frac{L^{2}}{\sin^{2}\theta}\bigg)-2an\cos\theta(1-2E^{2})\notag\\&&-\frac{4n\cos\theta E}{\sin^{2}\theta}\Big(n\cos\theta E+L\Big), \label{20}\\
R=R(r)&=&\big(E(a\chi+\Sigma)-aL\big)^{2}-\Delta\big(K+r^{2}+n^{2}+(L-aE)^{2}\big).\label{21}
\end{eqnarray}
The $\pm$ signs are independent from each other, but one must be consistent in that choice. The $+(-)$ sign corresponds to the outgoing(ingoing) geodesics. Clearly, the Carter constant $K$ vanishes for the equations of motion in the equatorial plane $\Big(\theta=$ {\Large $\frac{\pi}{2}$}$\Big)$. The radial equation of motion (\ref{19}) can also be written as
\begin{equation}\label{22}
\frac{1}{2}(u^{r})^{2}+V_{\text{eff}}(r,\theta)=\frac{1}{2}(E^{2}-1),
\end{equation}
with the effective potential
\begin{eqnarray}\label{23}
V_{\text{eff}}(r,\theta)&=&-\frac{2Mr+2n^{2}-a^{2}\sin^{2}\theta-Q^2+2an\cos\theta}{2\big(r^{2}+(n+a\cos^{2}\theta)^{2}\big)}
+\frac{1}{2\big(r^{2}+(n+a\cos^{2}\theta)^{2}\big)^{2}}\notag\\&&\times\bigg(L^{2}\big(r^2-2Mr-n^{2}+Q^{2}\big)+\big(r^2-2Mr-n^{2}+a^{2}+Q^{2}\big)\big(K-a\cos\theta(a\cos\theta+2n)\big)
\notag\\&&+\Big(a\cos\theta\big(a^{3}\cos^{3}\theta+2a\cos\theta(r^{2}+3n^{2}-2an\cos\theta)+4n(r^{2}+n^{2})\big)\notag\\&&-a^{2}\big(3n^{2}+r^{2}+2Mr-Q^{2}\big)\Big)E^{2}
+2aLE(2Mr+2n^2-Q^2)\bigg).
\end{eqnarray}
From Eqs. (\ref{19}) and (\ref{22}), we conclude that $V_{\text{eff}}(r,\theta)=$ {\Large $\frac{1}{2}$} $(E^2-1)-${\Large $\frac{R(r)}{2\Sigma^2}$}. Note that from Eqs. (\ref{18}) and (\ref{19}), for the allowed motion $\Theta\geq0$ and $R\geq0$ must be satisfied. Hence, the allowed and prohibited regions for the effective potential are given by $V_{\text{eff}}(r,\theta)\leq${\Large $\frac{1}{2}$} $(E^2-1)$ and $V_{\text{eff}}(r,\theta)>${\Large $\frac{1}{2}$} $\times(E^2-1)$, respectively. Also, $V_{\text{eff}}(r,\theta)\rightarrow0$ as $r\rightarrow\infty$.\\
In the equatorial plane, the effective potential is given by
\begin{eqnarray}\label{24}
V_{\text{eff}}\Big(r,\frac{\pi}{2}\Big)&=&-\frac{2Mr+2n^{2}-a^{2}-Q^2}{2\big(r^{2}+n^{2}\big)}
+\frac{1}{2\big(r^{2}+n^{2}\big)^{2}}\bigg(L^{2}\big(r^2-2Mr-n^{2}+Q^{2}\big)\notag\\&&-a^{2}\big(3n^{2}+r^{2}+2Mr-Q^{2}\big)E^{2}
+2aLE\big(2Mr+2n^2-Q^2\big)\bigg).
\end{eqnarray}
The function $R(r)$ can also be written in the form
\begin{eqnarray}\label{25}
R(r)&=&(E^{2}-1)r^{4}+2Mr^{3}+[(E^{2}-1)a^{2}-L^{2}+2E^{2}n^{2}-Q^{2}-K]r^{2}+2M[(L-aE)^{2}+n^{2}\notag\\&&+K]r+
E^{2}(3a^{2}n^{2}+n^{4}-a^{2}Q^{2})+L^{2}(n^{2}-Q^{2})+2aLE(-2n^{2}+Q^{2})-(n^{2}+K)\notag\\&&\times(a^{2}-n^{2}+Q^{2}).
\end{eqnarray}
Note that coefficient of the highest power of $r$ on the right-hand side is positive if $E>1$. Only in this case, the motion can be unbounded (infinite). For $E<1$, the motion is bounded (finite) i.e., the particle cannot reach the horizon(s) of the black hole. For $E=1$, the motion is marginally bounded i.e., the motion is either finite or infinite. In this case, the particle's motion depends on the black hole parameters and specific angular momentum for the allowed and prohibited regions of $R(r)$ and $\Theta(\theta)$ but the motion can be fully analysed by $R(r)$ or $V_{\text{eff}}\Big(r,$ {\Large $\frac{\pi}{2}$}$\Big)$ in the equatorial plane. The particle whose motion is bounded, unbounded and marginally bounded are respectively called bound, unbound and marginally bound particle. For bound and marginally bound particles, we have $V_{\text{eff}}(r,\theta)<0$ and $V_{\text{eff}}(r,\theta)\leq0$, respectively.
\\
We need to impose the condition $u^{t}>0$ along the geodesic. This is called the``forward-in-time" condition which shows that the time coordinate $t$ increases along the trajectory of the particle's motion. From Eq. (\ref{13}), this condition reduces to
\begin{equation}\label{26}
E\big[(\Sigma+a\chi)^{2}\sin^{2}\theta-\chi^{2}\Delta\big]>L\big[a(\Sigma+a\chi)\sin^{2}\theta-\chi\Delta\big].
\end{equation}
For $r\rightarrow r_{+}$, Eq. (\ref{26}) implies
\begin{equation}\label{27}
L\leq\frac{E\big[2(n^2+Mr_{+})-Q^{2}\big]}{a}.
\end{equation}
Here, we get the upper bound of the specific angular momentum at the outer horizon of the non-extremal KNTN black hole which is called the critical angular momentum and is denoted by $\hat{L}_{+}$ i.e.,
\begin{equation}\label{28}
\hat{L}_{+}=\frac{E\big[2(n^2+Mr_{+})-Q^{2}\big]}{a}.
\end{equation}
Similarly, the critical angular momentum at the inner horizon of the non-extremal KNTN black hole is given by
\begin{equation}\label{29}
\hat{L}_{-}=\frac{E\big[2(n^2+Mr_{-})-Q^{2}\big]}{a}.
\end{equation}
For the extremal KNTN black hole, we use $r_{+}=M$ in Eq. (\ref{28}), which gives the critical angular momentum at the horizon of the extremal KNTN black hole
\begin{equation}\label{30}
\hat{L}=\frac{E\big(2a^{2}+Q^{2}\big)}{a}.
\end{equation}
For $a=0$, Eqs. (\ref{28}), (\ref{29}) and (\ref{30}) become ill-defined, so we will assume $a\neq0$ throughout our work.
\section{Center of Mass Energy for two neutral particles}
 In this section, we will study the CME of the collision for two neutral particles with different rest masses falling freely from rest at infinity towards a KNTN black hole. Let us consider that these particles collide at some radial coordinate $r$ which are not restricted in the equatorial plane. The 4-momentum of the $i$th particle is given by
\begin{equation}\label{31}
P^{\mu}_{i}=m_{i}u^{\mu}_{i},
\end{equation}
where $i=1,2$ and $P^{\mu}_{i}$, $u^{\mu}_{i}$ and $m_{i}$ are respectively the 4-momentum, 4-velocity and rest mass (mass at rest at infinity) of the $i$th particle. The total 4-momentum of the two particles is
\begin{equation}\label{32}
P_{T}^{\mu}=P_{(1)}^{\mu}+P_{(2)}^{\mu}.
\end{equation}
Since the 4-momentum has zero spatial components in the center of mass frame, therefore the CME for the two particles is
\begin{equation}\label{33}
E_{\text{cm}}^{2}=-P_{T}^{\mu}P_{T\mu}=-(m_{1}u^{\mu}_{(1)}+m_{2}u^{\mu}_{(2)})(m_{1}u_{(1)\mu}+m_{2}u_{(2)\mu}).
\end{equation}
Simplifying and using $u^{\mu}_{(i)}u_{(i)\mu}=-1$ in Eq. (\ref{33}), we obtain
\begin{equation}\label{34}
\frac{E_{\text{cm}}}{\sqrt{2m_{1}m_{2}}}=\sqrt{\frac{(m_{1}-m_{2})^{2}}{2m_{1}m_{2}}+1-g_{\mu\nu}u_{(1)}^{\mu}u_{(2)}^{\nu}}.
\end{equation}
For the KNTN metric (\ref{1}), using Eqs. (\ref{13}), (\ref{14}), (\ref{18}) and (\ref{19}) into Eq. (\ref{34}), we get the CME of the collision
\begin{equation}\label{35}
\frac{E_{\text{cm}}}{\sqrt{2m_{1}m_{2}}}=\sqrt{\frac{(m_{1}-m_{2})^{2}}{2m_{1}m_{2}}+\frac{F(r, \theta)-G(r,\theta)-H(r,\theta)}{I(r,\theta)}},
\end{equation}
where $F(r, \theta)$, $G(r,\theta)$, $H(r,\theta)$ and $I(r,\theta)$ are given by
\begin{eqnarray}\label{36}
F(r,\theta)&=&\Delta\Sigma\sin^{2}\theta-(\Delta-a^{2}\sin^{2}\theta)L_{1}L_{2}+\big((\Sigma+a\chi)^{2}\sin^{2}\theta-\chi^{2}\Delta\big)E_{1}E_{2}\notag\\&&
+\big(\chi\Delta-a(\Sigma+a\chi)\sin^{2}\theta\big)\big(L_{1}E_{2}+L_{2}E_{1}\big),\notag\\
G(r,\theta)&=&\sin^{2}\theta\sqrt{R_{1}(r)R_{2}(r)},\notag\\
R_{i}(r)&=&\big[\big(r^{2}+n^{2}+a^{2}\big)E_{i}-aL_{i}\big]^{2}-\Delta\big[K_{i}+r^{2}+n^{2}+(L_{i}-aE_{i})^{2}\big],\notag\\
H(r,\theta)&=&\Delta\sin^{2}\theta\sqrt{\Theta_{1}(\theta)\Theta_{2}(\theta)},\\
\Theta_{i}(\theta)&=&K_{i}-2an\cos\theta\big(1-2E_{i}^{2}\big)-\cos^{2}\theta\bigg(\big(1-E_{i}^{2}\big)a^{2}+\frac{L_{i}^{2}}{\sin^{2}\theta}\bigg)
\notag\\&&-\frac{4n\cos\theta E_{i}}{\sin^{2}\theta}\bigg(n\cos\theta E_{i}+L_{i}\bigg),\notag\\
I(r,\theta)&=&\Delta\Sigma\sin^{2}\theta.\notag
\end{eqnarray}
Here, $E_{i}$, $L_{i}$ and $K_{i}$ are respectively the specific energy, specific angular momentum and Carter constant of the $i$th particle. Clearly, the CME (\ref{35}) is invariant under the interchange of the quantities $L_{1}\leftrightarrow L_{2}$, $E_{1}\leftrightarrow E_{2}$ and $m_{1}\leftrightarrow m_{2}$.
\subsection{Near-horizon collision of particles around the non-extremal KNTN black hole}
Let us discuss the properties of the CME (\ref{35}) as the particles approach the horizons $r_{+}$ and $r_{-}$ of the non-extremal KNTN black hole.
\subsubsection{Collision at the outer horizon}
The terms $F(r, \theta)-G(r,\theta)-H(r,\theta)$ and $I(r,\theta)$ of right-hand side of Eq. (\ref{35}) vanish at $r_{+}$. Using L'Hospital's rule and the identity $r_{+}^{2}-2Mr_{+}-n^{2}+a^{2}+Q^{2}=0$, the value of the CME at $r_{+}$ becomes
\begin{equation}\label{37}
\frac{E_{\text{cm}}}{\sqrt{2m_{1}m_{2}}}\bigg|_{r\rightarrow r_{+}}=\sqrt{\frac{(m_{1}-m_{2})^{2}}{2m_{1}m_{2}}+\frac{\partial_{r}F(r, \theta)-\partial_{r}G(r,\theta)-\partial_{r}H(r,\theta)}{\partial_{r}I(r,\theta)}}\Bigg|_{r\rightarrow r_{+}},
\end{equation}
where
\begin{eqnarray}\label{38}
\partial_{r}F(r,\theta)\big|_{r\rightarrow r_{+}}&=& 2\big(r_{+}-M\big)\big(r_{+}^{2}+(n+a\cos\theta)^{2}\big)\sin^{2}\theta-2(r_{+}-M)L_{1}L_{2}\notag\\&&
+\big[4r_{+}\big(r_{+}^{2}+n^{2}+a^{2}\big)\sin^{2}\theta-2(r_{+}-M)(a\sin^{2}\theta-2n\cos\theta)^{2}\big]E_{1}E_{2}\notag\\&&
-2\big(2n(r_{+}-M)\cos\theta+aM\sin^{2}\theta\big)\big(L_{1}E_{2}+L_{2}E_{1}\big),\notag\\
\partial_{r}G(r,\theta)\big|_{r\rightarrow r_{+}}&=&\bigg[\frac{\sin^{2}\theta}{2\sqrt{R_{1}(r)R_{2}(r)}}\bigg(R_{2}(r) \partial_{r}R_{1}(r)+R_{1}(r)\partial_{r} R_{2}(r)\bigg)\bigg]\bigg|_{r\rightarrow r_{+}},\\
\partial_{r}R_{i}(r)\big|_{r\rightarrow r_{+}}&=&4r_{+}\big[\big(r_{+}^{2}+n^{2}+a^{2}\big)E_{i}-aL_{i}\big]E_{i}-2\big(r_{+}-M\big)\big[K_{i}+r_{+}^{2}+n^{2}+(L_{i}-aE_{i})^{2}\big],\notag\\
\partial_{r}H(r,\theta)\big|_{r\rightarrow r_{+}}&=&2(r_{+}-M)\sin^{2}\theta\sqrt{\Theta_{1}(\theta)\Theta_{2}(\theta)},\notag\\
\partial_{r}I(r,\theta)\big|_{r\rightarrow r_{+}}&=&2(r_{+}-M)\big(r_{+}^{2}+(n+a\cos\theta)^{2}\big)\sin^{2}\theta.\notag
\end{eqnarray}
After much simplification, we get the CME at the outer horizon
\begin{eqnarray}\label{39}
\frac{E_{\text{cm}}}{2\sqrt{m_{1}m_{2}}}\bigg|_{r\rightarrow r_{+}}&=&\Bigg[\frac{(m_{1}-m_{2})^2}{4m_{1}m_{2}}+1+\frac{1}{4(\hat{L}_{+1}-L_{1})(\hat{L}_{+2}-L_{2})}\bigg[\big[
(\hat{L}_{+1}-L_{1})-(\hat{L}_{+2}-L_{2})\big]^{2}\notag\\&&+\frac{1}{r_{+}^{2}+(n+a\cos\theta)^{2}}\bigg(\frac{(r_{+}^{2}+n^{2})^{2}}{(r_{+}^{2}+n^{2}+a^{2})^{2}}
\big(L_{1}\hat{L}_{+2}-L_{2}\hat{L}_{+1}\big)^{2}+K_{2}(\hat{L}_{+1}-L_{1})^{2}\notag\\&&+K_{1}(\hat{L}_{+2}-L_{2})^{2}-a\cos\theta\big(2n+a\cos\theta\big)
\big[(\hat{L}_{+1}-L_{1})^{2}+(\hat{L}_{+2}-L_{2})^{2}\big]\bigg)\bigg]\notag\\&&-
\frac{1}{2\big(r_{+}^{2}+(n+a\cos\theta)^{2}\big)\sin^{2}\theta}\bigg[\cos^{2}\theta L_{1}L_{2}+
\frac{2an\cos\theta(L_{1}\hat{L}_{+2}+L_{2}\hat{L}_{+1})}{r_{+}^{2}+n^{2}+a^{2}}\notag\\&&+\frac{a^{2}\big[(a\sin^{2}\theta-2n\cos\theta)^{2}-a^{2}\sin^{2}\theta\big]}
{(r_{+}^{2}+n^{2}+a^{2})^{2}}\hat{L}_{+1}\hat{L}_{+2}+\sin^{2}\theta\sqrt{\Theta_{1}(\theta)\Theta_{2}(\theta)}\bigg]\Bigg]^{\frac{1}{2}},
\end{eqnarray}
where $\hat{L}_{+i}$ is the critical angular momentum for the $i$th particle, and can be written as $\hat{L}_{+i}=$ {\Large $\frac{E_{i}[2(n^{2}+Mr_{+})-Q^{2}]}{a}$}. The necessary condition to obtain an arbitrarily high CME is $L_{i}=\hat{L}_{+i}$. Choosing $E_{1}=E_{2}=E$, we get $\hat{L}_{+1}=\hat{L}_{+2}=\hat{L}_{+}=$ {\Large $\frac{E[2(n^{2}+Mr_{+})-Q^{2}]}{a}$}, and Eq. (\ref{39}) reduces to
\begin{eqnarray}\label{40}
\frac{E_{\text{cm}}}{2\sqrt{m_{1}m_{2}}}\bigg|_{r\rightarrow r_{+}}&=&\Bigg[\frac{(m_{1}-m_{2})^2}{4m_{1}m_{2}}+1+\frac{1}{4(\hat{L}_{+}-L_{1})(\hat{L}_{+}-L_{2})}\bigg[(L_{1}-L_{2})^{2}\notag\\&&
+\frac{1}{r_{+}^{2}+(n+a\cos\theta)^{2}}\bigg(\frac{(r_{+}^{2}+n^{2})^{2}}{a^{2}}
E^{2}\big(L_{1}-L_{2}\big)^{2}+K_{2}(\hat{L}_{+}-L_{1})^{2}\notag\\&&+K_{1}(\hat{L}_{+}-L_{2})^{2}-a\cos\theta\big(2n+a\cos\theta\big)
\big[(\hat{L}_{+}-L_{1})^{2}+(\hat{L}_{+}-L_{2})^{2}\big]\bigg)\bigg]\notag\\&&-
\frac{1}{2\big(r_{+}^{2}+(n+a\cos\theta)^{2}\big)\sin^{2}\theta}\bigg[\cos^{2}\theta L_{1}L_{2}
+2n\cos\theta E(L_{1}+L_{2})\notag\\&&+\big[(a\sin^{2}\theta-2n\cos\theta)^{2}-a^{2}\sin^{2}\theta\big]E^{2}
+\sin^{2}\theta\sqrt{\Theta_{1}(\theta)\Theta_{2}(\theta)}\bigg]\Bigg]^{\frac{1}{2}}.
\end{eqnarray}

Let us consider a marginally bound particle $(E=1)$ with the critical angular momentum $\hat{L}_{+}$. The conditions for the allowed region, $R(r)\geq0$ and $\Theta(\theta)\geq0$ give the upper and lower bounds for the Carter constant $K$ given below
\begin{equation}\label{41}
K^{(1)}_{\text{min}}\leq K\leq K^{(1)}_{\text{max}},
\end{equation}
where
\begin{eqnarray}
K^{(1)}_{\text{max}}&=&\frac{(r+r_{+})^{2}(r-r_{+})}{(r-r_{-})}-r^{2}-\bigg(\frac{n^{2}+r_{+}^{2}}{a}\bigg)^{2}-n^{2}, \label{42}\\
 K^{(1)}_{\text{min}}&=&\frac{\cos^{2}\theta}{a^{2}\sin^{2}\theta}\Big((a^{2}+n^{2}+r_{+}^{2})^{2}+4a^{2}n^{2}\Big)
 +\frac{4n\cos\theta}{a\sin^{2}\theta}(a^{2}+n^{2}+r_{+}^{2})-2an\cos\theta. \label{43}
\end{eqnarray}
Solving Eq. (\ref{41}), we find that the marginally bound particle with the critical angular momentum reaches the outer horizon of the non-extremal KNTN black hole if the following condition is satisfied
\begin{equation}\label{44}
A_{1}\cos^{3}\theta+B_{1}\cos^{2}\theta
+C_{1}\cos\theta+D_{1}\leq0~~~~\text{for~any}~~~~r\geq r_{+},
\end{equation}
where $A_{1}=2a^{3}n$, $B_{1}=a^{2}\Big($ {\Large $\frac{(r+r_{+})^{2}(r-r_{+})}{(r-r_{-})}$}$-r^{2}+5n^{2}+a^{2}+2r_{+}^{2}\Big)$,
$C_{1}=2an(a^{2}+2n^{2}+2r_{+}^{2})$ and $D_{1}=-a^{2}${\Large $\frac{(r+r_{+})^{2}(r-r_{+})}{(r-r_{-})}$}$+a^{2}r^{2}
+(n^{2}+r_{+}^{2})^{2}+a^{2}n^{2}$.

\begin{figure}[h!]
\centering
\includegraphics[width=10cm, height=8cm]{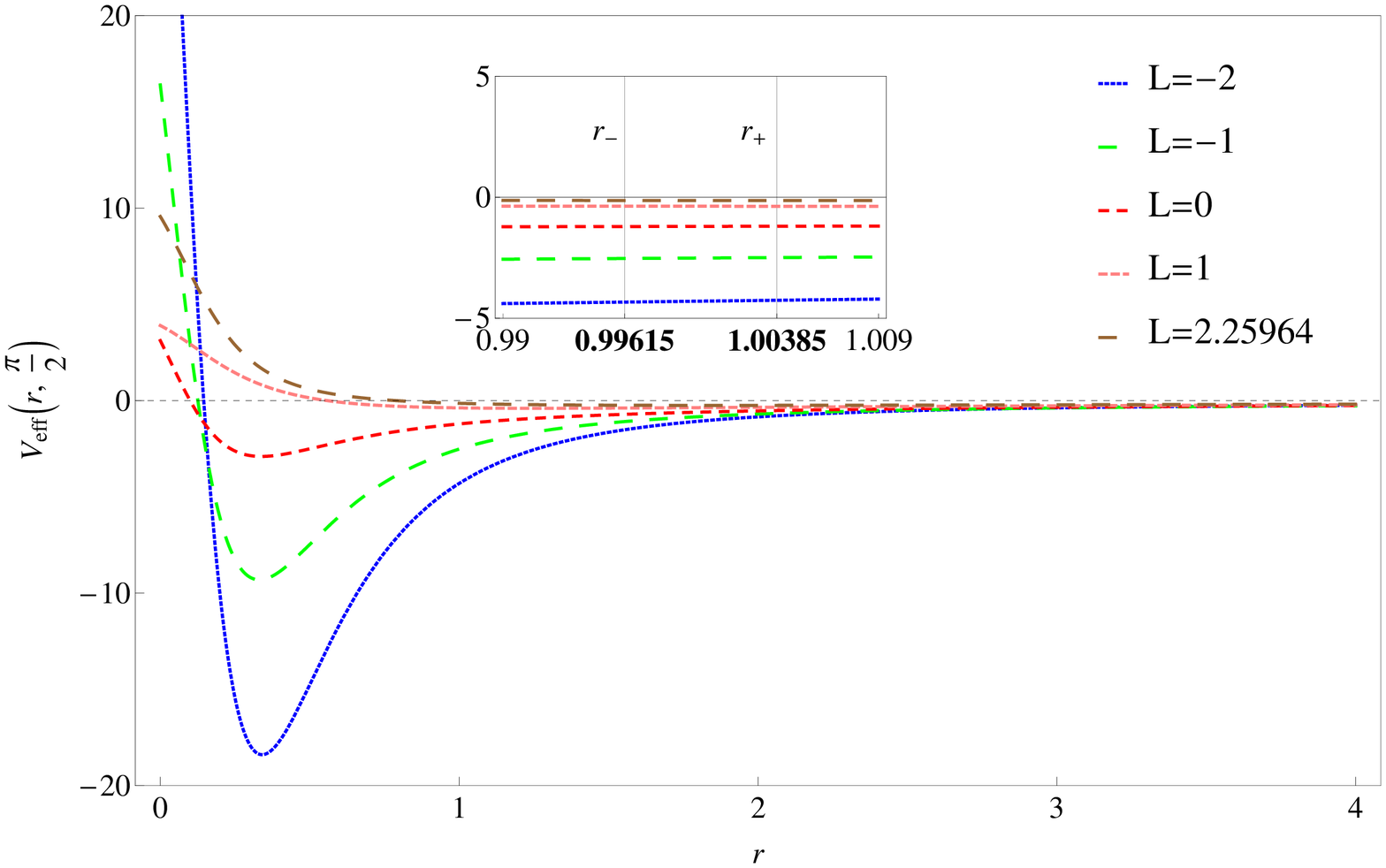}~~~~~~~~~~\\
\includegraphics[width=10cm, height=8cm]{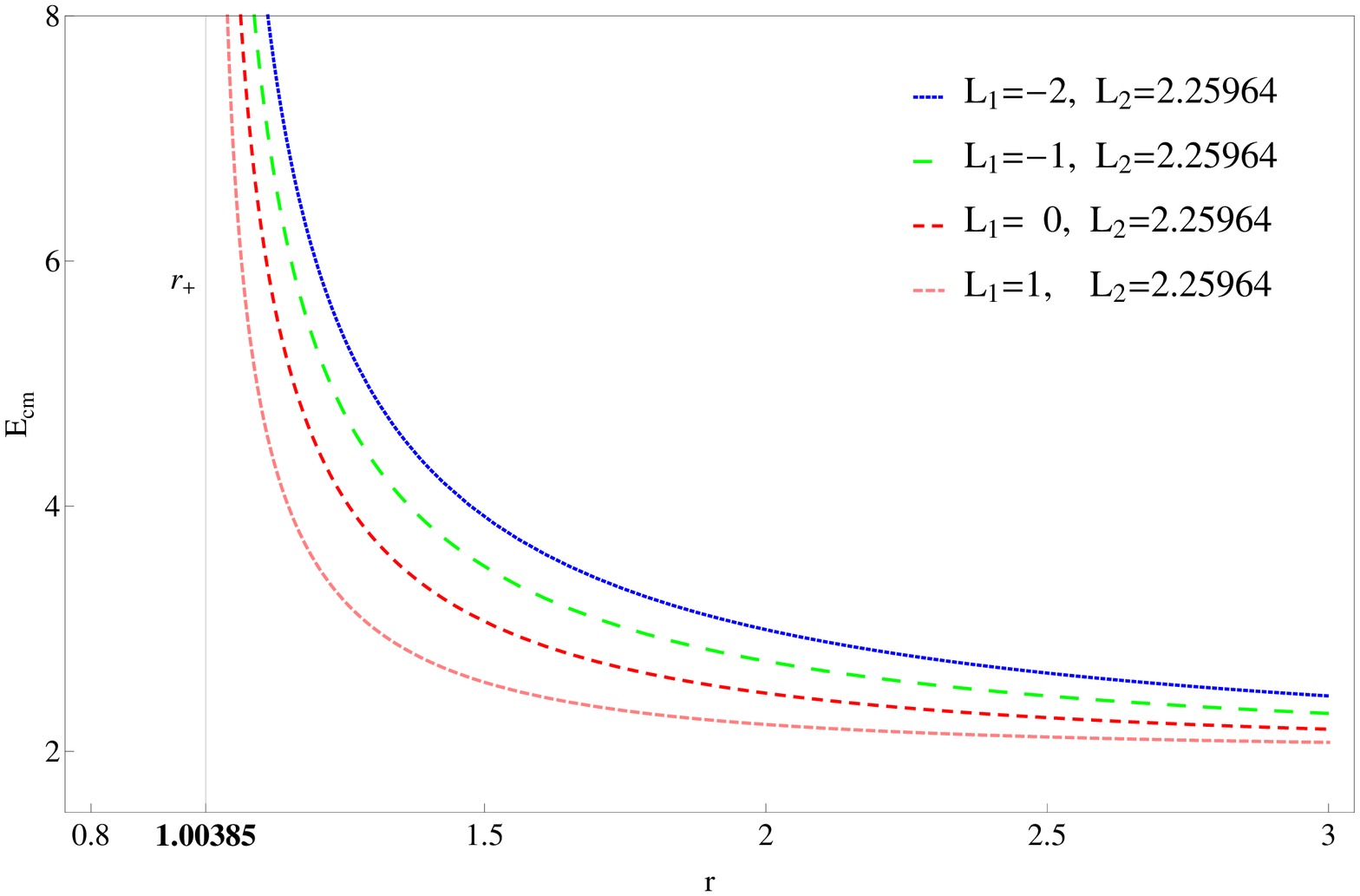}
\caption{The effective potential (top figure) and center of mass energy (bottom figure) for marginally bound particles in the equatorial plane of the non-extremal KNTN black hole. We set $M=1$, $m_{1}=m_{2}=1$, $a=0.8$, $n=0.4$ and $Q=0.7211$. Vertical lines identify the location of the inner and outer horizons of the black hole.}\label{KNTNf1}
\end{figure}

If one chooses $\theta=$ {\Large $\frac{\pi}{2}$}, the CME (\ref{39}) at the outer horizon of the non-extremal KNTN black hole reduces to
\begin{eqnarray}\label{45}
\frac{E_{\text{cm}}}{2\sqrt{m_{1}m_{2}}}\bigg|_{r\rightarrow r_{+}}&=&\bigg[\frac{(m_{1}-m_{2})^2}{4m_{1}m_{2}}+1+\frac{1}{4(\hat{L}_{+1}-L_{1})(\hat{L}_{+2}-L_{2})}\bigg(\big[
(\hat{L}_{+1}-L_{1})\notag\\&&-(\hat{L}_{+2}-L_{2})\big]^{2}+\frac{\big(L_{1}\hat{L}_{+2}-L_{2}\hat{L}_{+1}\big)^{2}}
{2M^{2}+2n^{2}-Q^{2}+2M\sqrt{M^{2}+n^{2}-a^{2}-Q^{2}}}\notag\\&&
-\frac{a^{2}\big(L_{1}\hat{L}_{+2}-L_{2}\hat{L}_{+1}\big)^{2}}{\big(2M^{2}+2n^{2}-Q^{2}+2M\sqrt{M^{2}+n^{2}-a^{2}-Q^{2}}\big)^{2}}\bigg)\bigg]^{\frac{1}{2}},
\end{eqnarray}
which is indeed finite for all values of $L_{1}$ and $L_{2}$ except when $L_{1}$ or $L_{2}$ is approximately equal to the critical angular momentum $\hat{L}_{+i}$, for which the neutral particles collide with an arbitrarily high CME. In the case of the same specific energies, the form of the CME (\ref{45}) at $r_{+}$  reads
\begin{eqnarray}\label{46}
\frac{E_{\text{cm}}}{2\sqrt{m_{1}m_{2}}}\bigg|_{r\rightarrow r_{+}}&=&\bigg[\frac{(m_{1}-m_{2})^2}{4m_{1}m_{2}}+1+\frac{1}{4(\hat{L}_{+}-L_{1})(\hat{L}_{+}-L_{2})}\bigg((L_{1}-L_{2})^{2}\notag\\&&
+\frac{2M^{2}+2n^{2}-a^{2}-Q^{2}+2M\sqrt{M^{2}+n^{2}-a^{2}-Q^{2}}}{a^{2}}
E^{2}\big(L_{1}-L_{2}\big)^{2}\bigg)\bigg]^{\frac{1}{2}}.
\end{eqnarray}
In Figure $\ref{KNTNf1}$, we plot the effective potential $V_{\text{eff}}\Big(r,$ {\Large $\frac{\pi}{2}$}$\Big)$ of marginally bound particles for $M=1,~a=0.8,~n=0.4,~Q=0.7211$ with different specific angular momenta $L=-2,~-1,~0,~1,~2.25964$ where $2.25964$ is the critical angular momentum $\hat{L}_{+}$. Clearly, the effective potential $V_{\text{eff}}\Big(r,$ {\Large $\frac{\pi}{2}$}$\Big)$ is negative when $r\geq r_{+}$, therefore the particles can reach the outer horizon. Vertical lines in the subplot represent the locations of the outer and inner horizons. We also plot the CME of the collision for $L_{1}=-2,~-1,~0,~1$ and $L_{2}=\hat{L}_{+}$. Clearly, the CME blows up at the outer horizon $r_{+}=1.00385$.
\subsubsection{Collision at the inner horizon}
Similarly the terms $F(r, \theta)-G(r,\theta)-H(r,\theta)$ and $I(r,\theta)$ of right-hand side of Eq. (\ref{35}) also vanish at $r_{-}$. Using L'Hospital's rule and by simplifying the calculation, we get the CME for the two neutral particles at the inner horizon
\begin{eqnarray}\label{47}
\frac{E_{\text{cm}}}{2\sqrt{m_{1}m_{2}}}\bigg|_{r\rightarrow r_{-}}&=&\Bigg[\frac{(m_{1}-m_{2})^2}{4m_{1}m_{2}}+1+\frac{1}{4(\hat{L}_{-1}-L_{1})(\hat{L}_{-2}-L_{2})}\bigg[\big[
(\hat{L}_{-1}-L_{1})-(\hat{L}_{-2}-L_{2})\big]^{2}\notag\\&&+\frac{1}{r_{-}^{2}+(n+a\cos\theta)^{2}}\bigg(\frac{(r_{-}^{2}+n^{2})^{2}}{(r_{-}^{2}+n^{2}+a^{2})^{2}}
\big(L_{1}\hat{L}_{-2}-L_{2}\hat{L}_{-1}\big)^{2}+K_{2}(\hat{L}_{-1}-L_{1})^{2}\notag\\&&+K_{1}(\hat{L}_{-2}-L_{2})^{2}-a\cos\theta\big(2n+a\cos\theta\big)
\big[(\hat{L}_{-1}-L_{1})^{2}+(\hat{L}_{-2}-L_{2})^{2}\big]\bigg)\bigg]\notag\\&&-
\frac{1}{2\big(r_{-}^{2}+(n+a\cos\theta)^{2}\big)\sin^{2}\theta}\bigg[\cos^{2}\theta L_{1}L_{2}+
\frac{2an\cos\theta(L_{1}\hat{L}_{-2}+L_{2}\hat{L}_{-1})}{r_{-}^{2}+n^{2}+a^{2}}\notag\\&&+\frac{a^{2}\big[(a\sin^{2}\theta-2n\cos\theta)^{2}-a^{2}\sin^{2}\theta\big]}
{(r_{-}^{2}+n^{2}+a^{2})^{2}}\hat{L}_{-1}\hat{L}_{-2}+\sin^{2}\theta\sqrt{\Theta_{1}(\theta)\Theta_{2}(\theta)}\bigg]\Bigg]^{\frac{1}{2}}.
\end{eqnarray}
This is the CME formula for the two neutral particles, where $\hat L_{-i}$ is the critical angular momentum at the inner horizon, which can be written as $\hat L_{-i}=$ {\Large $\frac{E_{i}[2(n^{2}+Mr_{-})-Q^{2}]}{a}$}.
An arbitrary high CME can be obtained by using the condition $L_{i}=\hat{L}_{-i}$ for either of the two particles. The critical angular momentum is same when both particles have the same specific energy and is given by $\hat{L}_{-1}=\hat{L}_{-2}=\hat{L}_{-}=$ {\Large $\frac{E[2(n^{2}+Mr_{-})-Q^{2}]}{a}$}, while the CME (\ref{47}) reduces to
\begin{eqnarray}\label{48}
\frac{E_{\text{cm}}}{2\sqrt{m_{1}m_{2}}}\bigg|_{r\rightarrow r_{-}}&=&\Bigg[\frac{(m_{1}-m_{2})^2}{4m_{1}m_{2}}+1+\frac{1}{4(\hat{L}_{-}-L_{1})(\hat{L}_{-}-L_{2})}\bigg[(L_{1}-L_{2})^{2}\notag\\&&
+\frac{1}{r_{-}^{2}+(n+a\cos\theta)^{2}}\bigg(\frac{(r_{-}^{2}+n^{2})^{2}}{a^{2}}
E^{2}\big(L_{1}-L_{2}\big)^{2}+K_{2}(\hat{L}_{-}-L_{1})^{2}\notag\\&&+K_{1}(\hat{L}_{-}-L_{2})^{2}-a\cos\theta\big(2n+a\cos\theta\big)
\big[(\hat{L}_{-}-L_{1})^{2}+(\hat{L}_{-}-L_{2})^{2}\big]\bigg)\bigg]\notag\\&&-
\frac{1}{2\big(r_{-}^{2}+(n+a\cos\theta)^{2}\big)\sin^{2}\theta}\bigg[\cos^{2}\theta L_{1}L_{2}
+2n\cos\theta E(L_{1}+L_{2})\notag\\&&+\big[(a\sin^{2}\theta-2n\cos\theta)^{2}-a^{2}\sin^{2}\theta\big]E^{2}
+\sin^{2}\theta\sqrt{\Theta_{1}(\theta)\Theta_{2}(\theta)}\bigg]\Bigg]^{\frac{1}{2}}.
\end{eqnarray}

Let us consider a marginally bound particle $(E=1)$ with the critical angular momentum $\hat{L}_{-}$. The conditions for the allowed region, $R(r)\geq0$ and $\Theta(\theta)\geq0$ give
\begin{eqnarray}\label{49}
K^{(2)}_{\text{min}}\leq K\leq K^{(2)}_{\text{max}},
\end{eqnarray}
where $K^{(2)}_{\text{min}}$ and $K^{(2)}_{\text{max}}$ are given by
\begin{eqnarray}
K^{(2)}_{\text{max}}&=&\frac{(r+r_{-})^{2}(r-r_{-})}{(r-r_{+})}-r^{2}-\bigg(\frac{n^{2}+r_{-}^{2}}{a}\bigg)^{2}-n^{2}, \label{50}\\
 K^{(2)}_{\text{min}}&=&\frac{\cos^{2}\theta}{a^{2}\sin^{2}\theta}\Big((a^{2}+n^{2}+r_{-}^{2})^{2}+4a^{2}n^{2}\Big)
 +\frac{4n\cos\theta}{a\sin^{2}\theta}(a^{2}+n^{2}+r_{-}^{2})-2an\cos\theta. \label{51}
\end{eqnarray}
The inequality (\ref{49}) gives the upper and lower bounds for the Carter constant $K$. By Eq. (\ref{49}), one can say that the marginally bound particle with the critical angular momentum reaches the inner horizon of the non-extremal KNTN black hole if the following condition is satisfied
\begin{equation}\label{52}
A_{2}\cos^{3}\theta+B_{2}\cos^{2}\theta
+C_{2}\cos\theta+D_{2}\leq0~~~~\text{for~any}~~~~r\geq r_{-},
\end{equation}
where $A_{2}=2a^{3}n$, $B_{2}=a^{2}\Big(${\Large $\frac{(r+r_{-})^{2}(r-r_{-})}{(r-r_{+})}$}$-r^{2}+5n^{2}+a^{2}+2r_{-}^{2}\Big)$, $C_{2}=2an(a^{2}+2n^{2}+2r_{-}^{2})$ and $D_{2}=-a^{2}${\Large $\frac{(r+r_{-})^{2}(r-r_{-})}{(r-r_{+})}$}$+a^{2}r^{2}
+(n^{2}+r_{-}^{2})^{2}+a^{2}n^{2}$.

\begin{figure}[h!]
\centering
\includegraphics[width=10cm, height=8cm]{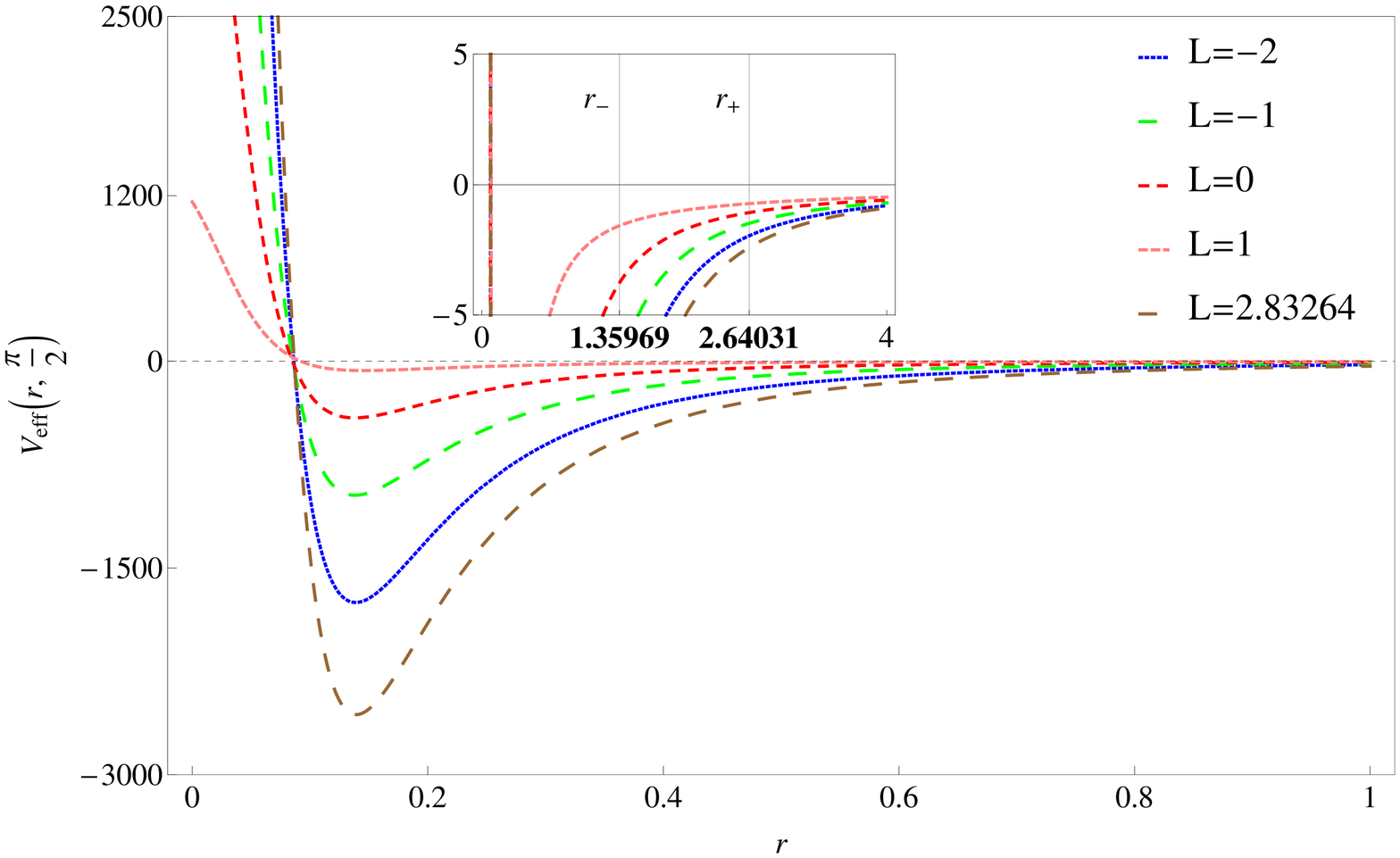}~~~~~~~~~~\\
\includegraphics[width=10cm, height=8cm]{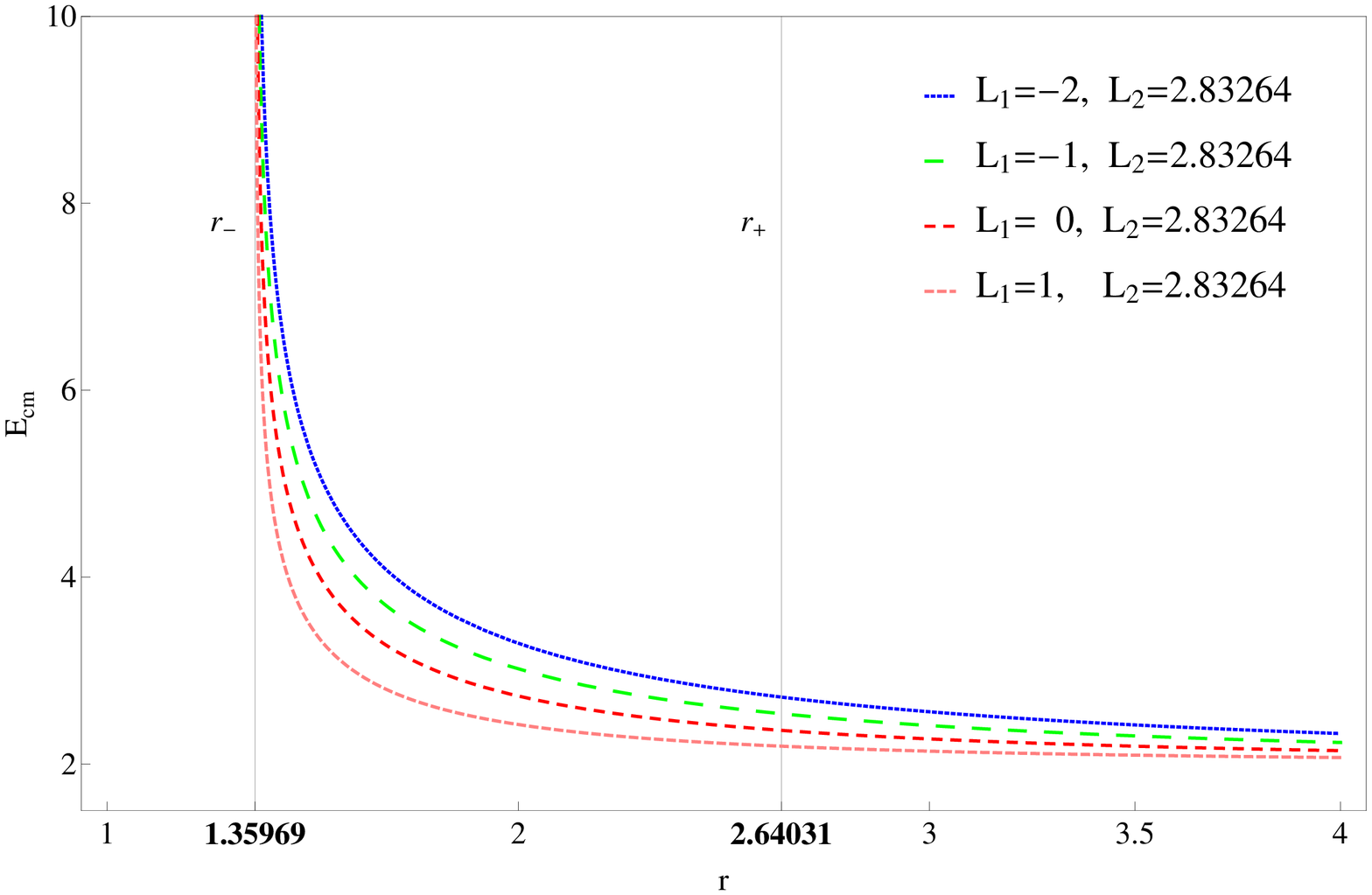}
\caption{The effective potential (top figure) and center of mass energy (bottom figure) for marginally bound particles in the equatorial plane of the non-extremal KNTN black hole. We set $M=2$, $m_{1}=m_{2}=1$, $a=1.8$, $n=0.1$, and $Q=0.6$. Vertical lines identify the location of the inner and outer horizons of the black hole.}\label{KNTNf2}
\end{figure}

In the equatorial plane, Eq. (\ref{47}) at the inner horizon takes the following form
\begin{eqnarray}\label{53}
\frac{E_{\text{cm}}}{2\sqrt{m_{1}m_{2}}}\bigg|_{r\rightarrow r_{-}}&=&\bigg[\frac{(m_{1}-m_{2})^2}{4m_{1}m_{2}}+1+\frac{1}{4(\hat{L}_{-1}-L_{1})(\hat{L}_{-2}-L_{2})}\bigg(\big[
(\hat{L}_{-1}-L_{1})\notag\\&&-(\hat{L}_{-2}-L_{2})\big]^{2}+\frac{\big(L_{1}\hat{L}_{-2}-L_{2}\hat{L}_{-1}\big)^{2}}
{2M^{2}+2n^{2}-Q^{2}-2M\sqrt{M^{2}+n^{2}-a^{2}-Q^{2}}}\notag\\&&
-\frac{a^{2}\big(L_{1}\hat{L}_{-2}-L_{2}\hat{L}_{-1}\big)^{2}}{\big(2M^{2}+2n^{2}-Q^{2}-2M\sqrt{M^{2}+n^{2}-a^{2}-Q^{2}}\big)^{2}}\bigg)\bigg]^{\frac{1}{2}}.
\end{eqnarray}
Clearly, the CME is finite for all values of $L_{1}$ and $L_{2}$ except when $L_{1}$ or $L_{2}$ is approximately equal to the critical angular momentum. For $E_{1}=E_{2}=E$, Eq. (\ref{53}) gives
\begin{eqnarray}\label{54}
\frac{E_{\text{cm}}}{2\sqrt{m_{1}m_{2}}}\bigg|_{r\rightarrow r_{-}}&=&\bigg[\frac{(m_{1}-m_{2})^2}{4m_{1}m_{2}}+1+\frac{1}{4(\hat{L}_{-}-L_{1})(\hat{L}_{-}-L_{2})}\bigg((L_{1}-L_{2})^{2}\notag\\&&
+\frac{2M^{2}+2n^{2}-a^{2}-Q^{2}-2M\sqrt{M^{2}+n^{2}-a^{2}-Q^{2}}}{a^{2}}
E^{2}\big(L_{1}-L_{2}\big)^{2}\bigg)\bigg]^{\frac{1}{2}}.
\end{eqnarray}
We plot the effective potential $V_{\text{eff}}\Big(r,$ {\Large $\frac{\pi}{2}$}$\Big)$ of marginally bound particles in Figure $\ref{KNTNf2}$ for $M=2,~a=0.6,~n=0.1,~Q=0.6$ with different specific angular momenta $L=-2,~-1,~0,~1,~\hat{L}_{-}$ where $\hat{L}_{-}=2.83264$. Clearly, the effective potential $V_{\text{eff}}\Big(r,$ {\Large $\frac{\pi}{2}$}$\Big)$ is negative for $r\geq r_{-}$, so the particles can reach the inner horizon after crossing outer horizon. The subplot shows the behaviour of $V_{\text{eff}}\Big(r,$ {\Large $\frac{\pi}{2}$}$\Big)$ near the horizons and identify the location of the outer and inner horizons. We also plot the CME of the collision for $L_{1}=-2,~-1,~0,~1$ and $L_{2}=\hat{L}_{-}$. The CME is finite at the outer horizon and blows up at the inner horizon $r_{-}=1.35969$.
\subsection{Near-horizon collision of particles around the extremal KNTN black hole}
Let us study the properties of the CME (\ref{35}) as the particles approach the horizon of the extremal KNTN black hole. In the case of the extremal KNTN black hole, the NUT charge $n$, mass $M$, rotating parameter $a$ and charge $Q$ satisfies the relation $n^{2 }=a^{2}+Q^{2}-M^{2}$. Using this relation in Eq. (\ref{39}), we obtain
\begin{eqnarray}\label{55}
\frac{E_{\text{cm}}}{2\sqrt{m_{1}m_{2}}}\bigg|_{r\rightarrow M}&=&\Bigg[\frac{(m_{1}-m_{2})^2}{4m_{1}m_{2}}+1+\frac{1}{4(\hat{L}_{1}-L_{1})(\hat{L}_{2}-L_{2})}\bigg[\big[
(\hat{L}_{1}-L_{1})-(\hat{L}_{2}-L_{2})\big]^{2}\notag\\&&+\frac{1}{M^{2}+(n+a\cos\theta)^{2}}\bigg(\frac{(a^{2}+Q^{2})^{2}}{(2a^{2}+Q^{2})^{2}}
\big(L_{1}\hat{L}_{2}-L_{2}\hat{L}_{1}\big)^{2}+K_{2}(\hat{L}_{1}-L_{1})^{2}\notag\\&&+K_{1}(\hat{L}_{2}-L_{2})^{2}-a\cos\theta\big(2n+a\cos\theta\big)
\big[(\hat{L}_{1}-L_{1})^{2}+(\hat{L}_{2}-L_{2})^{2}\big]\bigg)\bigg]\notag\\&&-
\frac{1}{2\big(M^{2}+(n+a\cos\theta)^{2}\big)\sin^{2}\theta}\bigg[\cos^{2}\theta L_{1}L_{2}+
\frac{2an\cos\theta(L_{1}\hat{L}_{2}+L_{2}\hat{L}_{1})}{2a^{2}+Q^{2}}\notag\\&&+\frac{a^{2}\big[(a\sin^{2}\theta-2n\cos\theta)^{2}-a^{2}\sin^{2}\theta\big]}
{(2a^{2}+Q^{2})^{2}}\hat{L}_{1}\hat{L}_{2}+\sin^{2}\theta\sqrt{\Theta_{1}(\theta)\Theta_{2}(\theta)}\bigg]\Bigg]^{\frac{1}{2}}.
\end{eqnarray}
Eq. (\ref{55}) is the CME for the two neutral particles at the horizon of the extremal KNTN black hole. The critical angular momentum at the horizon is given by $\hat{L}_{i}=$ {\Large $\frac{E_{i}(2a^{2}+Q^{2})}{a}$}, for the $i$th particle. The necessary condition for obtaining an arbitrarily high CME is $L_{i}=\hat{L}_{i}$ for either of the two particles. For $E_{1}=E_{2}=E$, we get the same critical angular momentum i.e., $\hat{L}_{1}=\hat{L}_{2}=\hat{L}=$ {\Large $\frac{E(2a^{2}+Q^{2})}{a}$}, and Eq. (\ref{55}) takes the form
\begin{eqnarray}\label{56}
\frac{E_{\text{cm}}}{2\sqrt{m_{1}m_{2}}}\bigg|_{r\rightarrow M}&=&\Bigg[\frac{(m_{1}-m_{2})^2}{4m_{1}m_{2}}+1+\frac{1}{4(\hat{L}-L_{1})(\hat{L}-L_{2})}\bigg[(L_{1}-L_{2})^{2}\notag\\&&
+\frac{1}{M^{2}+(n+a\cos\theta)^{2}}\bigg(\frac{(a^{2}+Q^{2})^{2}}{a^{2}}
E^{2}\big(L_{1}-L_{2}\big)^{2}+K_{2}(\hat{L}-L_{1})^{2}\notag\\&&+K_{1}(\hat{L}-L_{2})^{2}-a\cos\theta\big(2n+a\cos\theta\big)
\big[(\hat{L}-L_{1})^{2}+(\hat{L}-L_{2})^{2}\big]\bigg)\bigg]\notag\\&&-
\frac{1}{2\big(M^{2}+(n+a\cos\theta)^{2}\big)\sin^{2}\theta}\bigg[\cos^{2}\theta L_{1}L_{2}+
2n\cos\theta E(L_{1}+L_{2})\notag\\&&+\big[(a\sin^{2}\theta-2n\cos\theta)^{2}-a^{2}\sin^{2}\theta\big]E^{2}
+\sin^{2}\theta\sqrt{\Theta_{1}(\theta)\Theta_{2}(\theta)}\bigg]\Bigg]^{\frac{1}{2}}.
\end{eqnarray}

Let us consider a marginally bound particle $(E=1)$ with the critical angular momentum $\hat{L}$. The inequality (\ref{41}) reduces to
\begin{equation}\label{57}
K^{(3)}_{\text{min}} \leq K\leq K^{(3)}_{\text{max}},
\end{equation}
where
\begin{eqnarray}
K^{(3)}_{\text{max}}&=&2Mr+M^{2}-\bigg(\frac{n^{2}+M^{2}}{a}\bigg)^{2}-n^{2},\label{58}\\
K^{(3)}_{\text{min}}&=&\frac{\cos^{2}\theta}{a^{2}\sin^{2}\theta}\Big((a^{2}+n^{2}+M^{2})^{2}+4a^{2}n^{2}\Big)
 +\frac{4n\cos\theta}{a\sin^{2}\theta}(a^{2}+n^{2}+M^{2})-2an\cos\theta. \label{59}
\end{eqnarray}
Thus for the marginally bound particle with the critical angular momentum to reach the horizon of the extremal KNTN black hole, the following condition must be satisfied
\begin{equation}\label{60}
A_{3}\cos^{3}\theta+B_{3}\cos^{2}\theta+C_{3}\cos\theta+D_{3}\leq0~~~~~~~~~~~~~~~~~~~\text{for~any}~~~~~~r\geq M,
\end{equation}
where $A_{3}=2a^{3}n$,~~ $B_{3}=a^{2}(2Mr+3M^{2}+5n^{2}+a^{2})$, ~~$C_{3}=2an(a^{2}+2n^{2}+2M^{2})$ and \\$D_{3}=-2a^{2}rM-a^{2}M^{2}+(n^2+M^{2})^{2}+a^{2}n^{2}$.

Further, if the collision occurs in the equatorial plane, the CME (\ref{55}) at the horizon of the extremal KNTN black hole reduces to
\begin{eqnarray}\label{61}
\frac{E_{\text{cm}}}{2\sqrt{m_{1}m_{2}}}\bigg|_{r\rightarrow M}&=&\bigg[\frac{(m_{1}-m_{2})^2}{4m_{1}m_{2}}+1+\frac{1}{4(\hat{L}_{1}-L_{1})(\hat{L}_{2}-L_{2})}\bigg(\big[
(\hat{L}_{1}-L_{1})-(\hat{L}_{2}-L_{2})\big]^{2}\notag\\&&+\frac{a^{2}+Q^{2}}{(2a^{2}+Q^{2})^{2}}
\big(L_{1}\hat{L}_{2}-L_{2}\hat{L}_{1}\big)^{2}\bigg)\bigg]^{\frac{1}{2}},
\end{eqnarray}
which is indeed finite for all values of $L_{1}$ and $L_{2}$ except when $L_{1}$ or $L_{2}$ approaches the critical angular momentum, for which the CME is arbitrarily high. When the electric charge $Q$ vanishes, Eq. (\ref{61}) gives the result for the extremal Kerr-Taub-NUT black hole as obtained in Ref. \cite{17}. When the specific energy of both the particles are exactly alike, then (\ref{61}) becomes
\begin{eqnarray}\label{62}
\frac{E_{\text{cm}}}{2\sqrt{m_{1}m_{2}}}\bigg|_{r\rightarrow M}&=&\bigg[\frac{(m_{1}-m_{2})^2}{4m_{1}m_{2}}+1+\frac{1}{4(\hat{L}-L_{1})(\hat{L}-L_{2})}\bigg((L_{1}-L_{2})^{2}\notag\\&&
+\frac{a^{2}+Q^{2}}{a^{2}}
E^{2}\big(L_{1}-L_{2}\big)^{2}\bigg)\bigg]^{\frac{1}{2}}.
\end{eqnarray}
There must exist intervals for the spin parameter $a$, NUT charge $n$ and electric charge $Q$ to ensure that the marginally bound particles with the critical angular momentum $\hat{L}$ reach the horizon of the extremal KNTN black hole and collide at the horizon. Since the motion of the particle in the equatorial plane can be fully analysed by the effective potential $V_{\text{eff}}\Big(r,${\Large $\frac{\pi}{2}$}$\Big)$, so with the help of the effective potential given in Eq. (\ref{24}), we can determine intervals of $a$ and $n$ corresponding to different values of $Q$. The effective potential for the marginally bound particle with the critical angular momentum $\hat{L}$ is given by
\begin{equation}\label{63}
V_{\text{eff}}\Big(r,\frac{\pi}{2}\Big)=-\frac{(r-M)^{2}\big(Mr-a^{2}+M^{2}-\frac{3Q^{2}}{2}-\frac{Q^{4}}{2a^{2}}\big)}
{(r^{2}+a^{2}+Q^{2}-M^{2})^{2}}.
\end{equation}
Here, the condition for the particle falling freely from rest at infinity to reach the horizon can be expressed as
\begin{equation}\label{64}
V_{\text{eff}}\Big(r,\frac{\pi}{2}\Big)\leq0~~~\text{for~any}~~~r\geq M,
\end{equation}
which is equivalent to
\begin{equation}\label{65}
Mr-a^{2}+M^{2}-\frac{3Q^{2}}{2}-\frac{Q^{4}}{2a^{2}}\geq0~~~\text{for~any}~~~r\geq M.
\end{equation}
Combining with the condition $0\leq n^{2}=a^{2}+Q^{2}-M^{2}$ and set $M=1$, we get intervals for $a$ and $n$ for different values of $Q$ as shown in Table ~\ref{KNTNtab:1}. Note that, there are two different intervals for the spin parameter $a$ corresponding to the co-rotating and counter-rotating orbits. For $Q=0$, we get intervals for $a$ and $n$ as discussed earlier in Ref. \cite{17}. With the increase of $Q$, the intervals for $a$ and $n$ become narrow.\\
The maximum and minimum value of $L$ can be obtained by the conditions
\begin{equation}\label{66}
V_{\text{eff}}\Big(r,\frac{\pi}{2}\Big)=0,~~~~\partial_{r}V_{\text{eff}}\Big(r,\frac{\pi}{2}\Big)=0.
\end{equation}
Then the interval $L\in[L_{\text{min}}, L_{\text{max}}]$ can be determined from it. The intervals for the specific angular momentum for different values of $a$ and $Q$ are shown in Table ~\ref{KNTNtab:2}. Note that, with the increase of $a$ and $Q$, the interval $L\in[L_{\text{min}}, L_{\text{max}}]$ becomes wider.
\begin{table}[h!]
\caption{The intervals for the spin parameter $a$ and NUT charge $n$ with different electric charge $Q$ for the extremal KNTN black hole.~~~~~~~~~~~~~~~~~~~~~~~~~~~~~~~~~~~~~~~~~~~~~~~~~~~~~~~~~~~~~~~~~~~~~~~~~~~~~~~~~~~~} \label{KNTNtab:1}
\begin{tabular}{|C{1.5cm}|C{9.5cm}|C{5cm}|}
 \hline
$Q$ & $a$ & $n$  \\
\hline\hline
$0$  & $\big[1, ~\sqrt{2}\big]$,~~~~~~~~~~~~~~~ $\big[-\sqrt{2}, ~-1\big]$ & $\big[-1, ~1\big]$  \\ \hline
$0.1$ & $\big[0.99499, ~1.40889\big]$,~~~ $\big[-1.40889, ~-0.99499\big]$ & $\big[-0.99748, ~0.99748\big]$  \\ \hline
$0.2$ & $\big[0.97980, ~1.39269\big]$,~~~ $\big[-1.39269, ~-0.97980\big]$ & $\big[-0.98974, ~0.98974\big]$   \\ \hline
$0.3$ & $\big[0.95394, ~1.36485\big]$,~~~ $\big[-1.36485, ~-0.95394\big]$ & $\big[-0.97612, ~0.97612\big]$  \\ \hline
$0.4$ & $\big[0.91652, ~1.32389\big]$,~~~ $\big[-1.32389, ~-0.91652\big]$ & $\big[-0.95534, ~0.95534\big]$   \\ \hline
\end{tabular}
\end{table}
\begin{table}[h!]
\caption {The interval $L\in[L_{\text{min}}, L_{\text{max}}]$ with different spin parameter $a$ and electric charge $Q$ for the extremal KNTN black hole.} \label{KNTNtab:2}
\begin{tabular}{|C{0.8cm}|C{3.75cm}|C{3.75cm}|C{3.75cm}|C{3.75cm}|}
 \hline
$Q$ & $a=1$ & $a=1.1$ & $a=1.2$ & $a=1.3$ \\
\hline\hline
$0$  & $\big[-4.82843, ~2\big]$ & $\big[-5.02685, ~2.20171\big]$ & $\big[-5.2224, ~2.34962\big]$ & $\big[-5.41546, ~2.58869\big]$   \\ \hline
$0.1$ & $\big[-4.83135, ~2.01\big]$ & $\big[-5.02955, ~2.06517\big]$ & $\big[-5.22492, ~2.36131\big]$ & $\big[-5.41782, ~2.59755\big]$   \\ \hline
$0.2$ & $\big[-4.84011, ~2.04\big]$ & $\big[-5.03766, ~2.23636\big]$ & $\big[-5.23246, ~2.39551\big]$ & $\big[-5.42488, ~2.62375\big]$   \\ \hline
$0.3$ & $\big[-4.85463, ~1.82908\big]$ & $\big[-5.05111, ~2.19607\big]$ & $\big[-5.24499, ~2.44978\big]$ & $\big[-5.4366, ~2.67045\big]$   \\ \hline
$0.4$ & $\big[-4.87481, ~2.16\big]$ & $\big[-5.06982, ~2.34748\big]$ & $\big[-5.26243, ~2.53484\big]$ & $\big[-5.45292, ~2.72286\big]$   \\ \hline
\end{tabular}
\end{table}

\begin{figure}[h!]
\includegraphics[width=10cm, height=8cm]{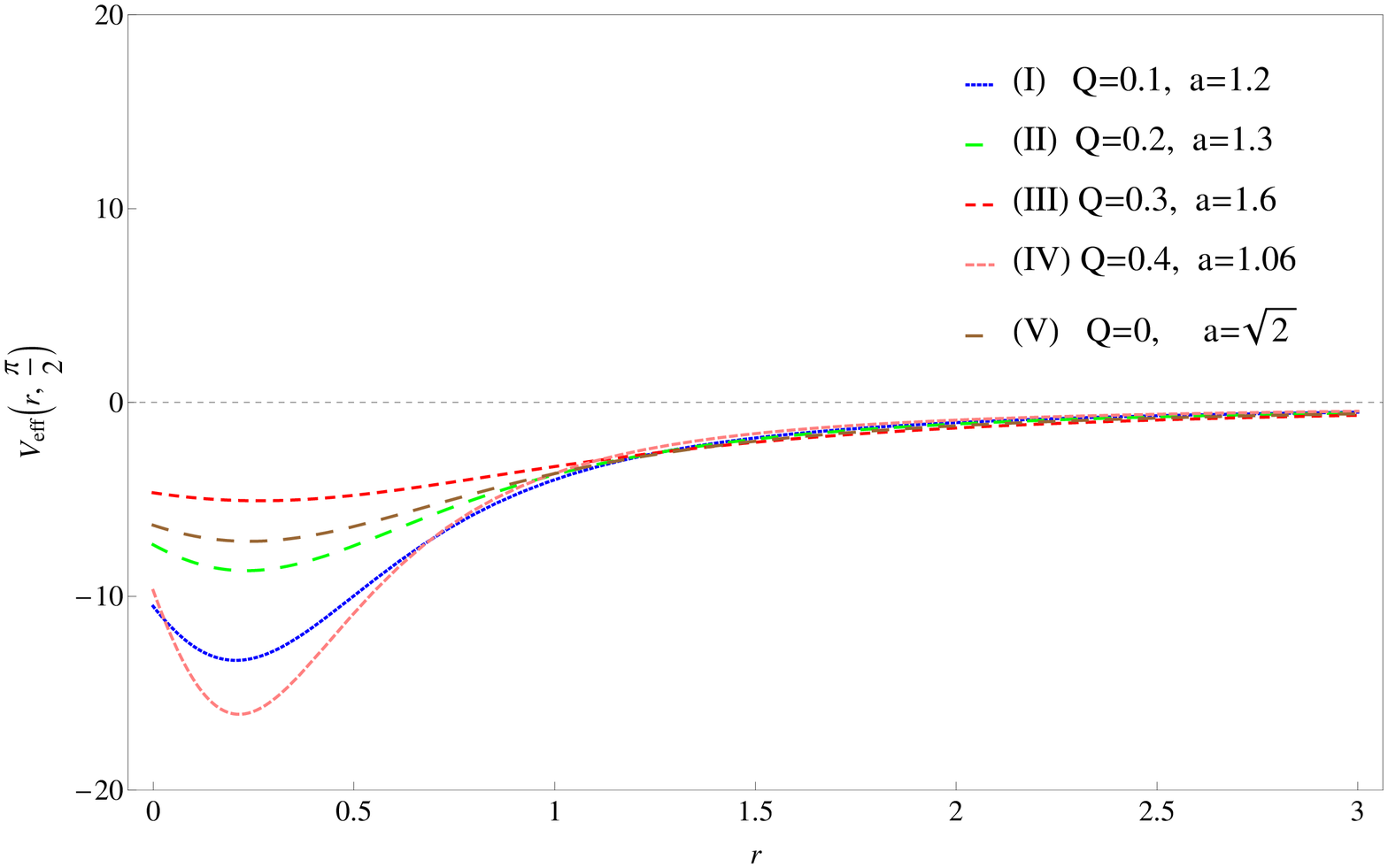}\\
\includegraphics[width=10cm, height=8cm]{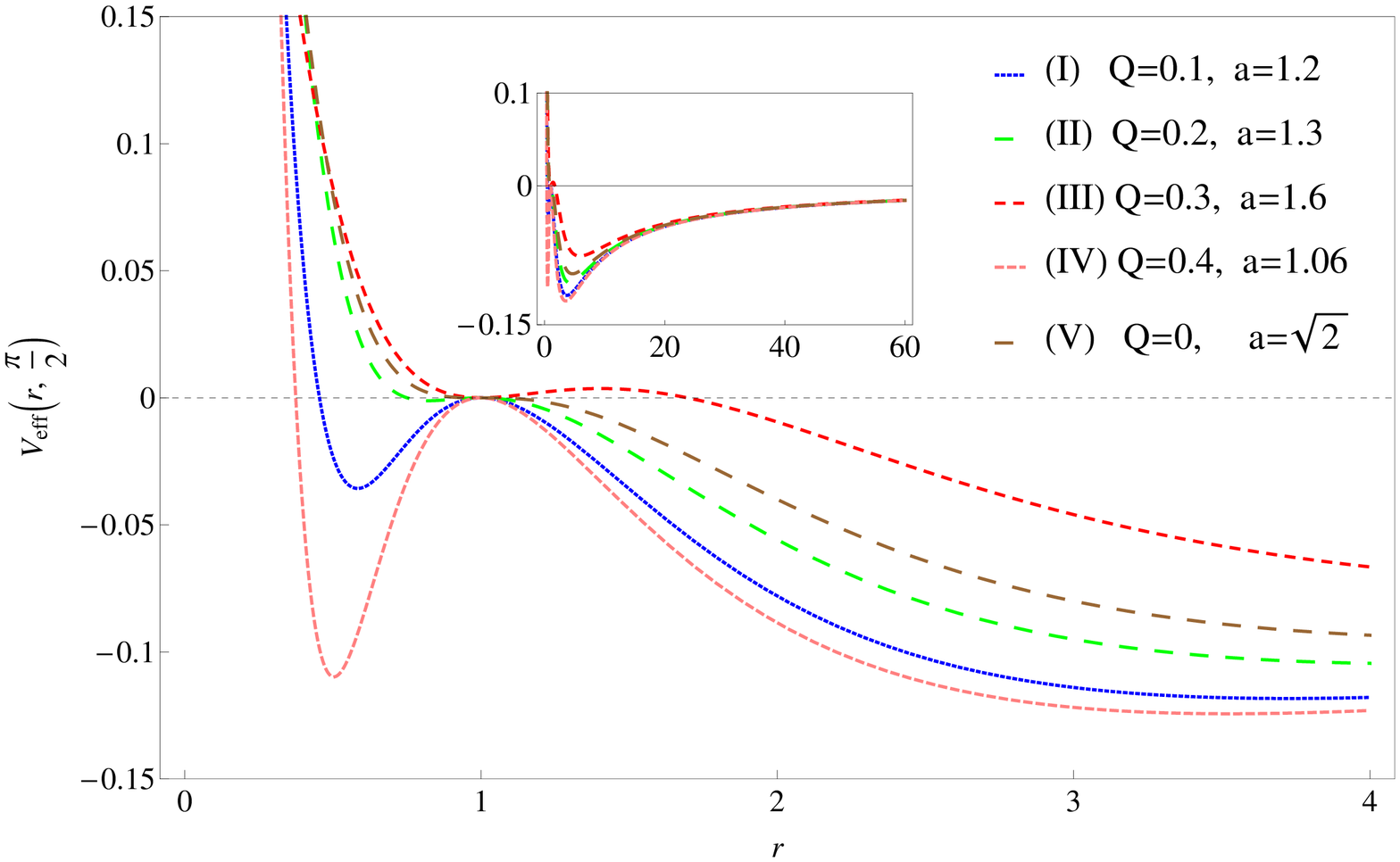}~~~~
\caption{The effective potential for marginally bound particles in the equatorial plane of the extremal KNTN black hole for $M=1$. Here, $L=-1$ in the top figure and $L=\hat{L}$ in the bottom figure. The horizon is fixed at $r_{+}=r_{-}=1$.}\label{KNTNf3}
\end{figure}

\begin{figure}[h!]
\centering
\includegraphics[width=10cm, height=8cm]{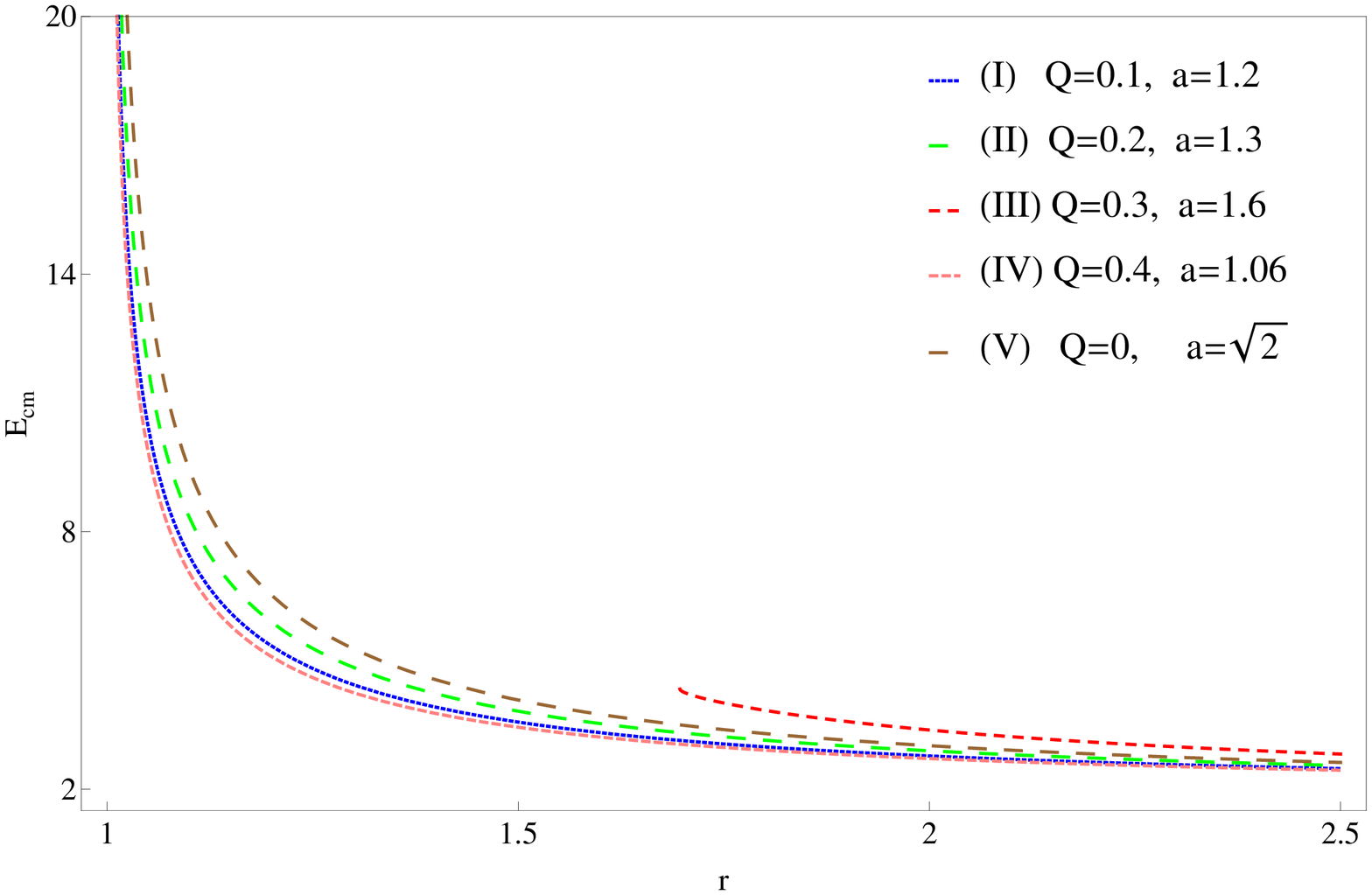}
\caption{The center of mass energy for marginally bound particles in the equatorial plane of the extremal KNTN black hole. We set $M=1$ and $m_{1}=m_{2}=1$. Here, $L_{1}=-1$ and $L_{2}=\hat{L}$. The horizon is at $r_{+}=r_{-}=1$.}\label{KNTNf4}
\end{figure}

We plot the effective potential $V_{\text{eff}}\Big(r,$ {\Large $\frac{\pi}{2}$}$\Big)$ in Figure $\ref{KNTNf3}$ for $L=-1$ and $L=\hat{L}$ in the top and bottom plots, respectively. Clearly, for case (III) $Q=0.3,~a=1.6$, the effective potential $V_{\text{eff}}\Big(r,$ {\Large $\frac{\pi}{2}$}$\Big)$ is non-positive for $L=-1$ but positive near the horizon $r_{+}=r_{-}=1$ for $L=\hat{L}$, so the particle cannot reach the horizon in this case for $L=\hat{L}$. For cases (I) $Q=0.1,~a=1.2$, (II) $Q=0.2,~a=1.3$, (IV) $Q=0.4,~a=1.06$ and (V) $Q=0,~a=\sqrt{2}$, $V_{\text{eff}}\Big(r,$ {\Large $\frac{\pi}{2}$}$\Big)\leq0$ when $r\geq M=1$ for the both specific angular momenta. Hence, the particle can reach the horizon in all the four cases for $L=-1$ and $L=\hat{L}$. We also plot the CME of the collision in Figure $\ref{KNTNf4}$ for $L_{1}=-1$ and $L_{2}=\hat{L}$. For the case (III) $Q=0.3,~a=1.6$, $a$ does not belong to $\big[0.95394, ~1.36485\big]$, the CME only exists for $r\geq1.69668$. This is because the collision for the two marginally bound particle with $L_{1}=-1$ and $L_{2}=\hat{L}$ cannot take place at $r<1.69668$. For the case (I), (II), (IV) and (V), the CME is divergent at the horizon $r_{+}=r_{-}=1$.
\section{Conclusion}
In this paper, we have studied the CME of the collision for two neutral particles with different
rest masses falling freely from rest at infinity in the background of a KNTN background. Further, we have discussed the CME when the collision takes place near the horizon(s) of an extremal and non-extremal KNTN black hole. We have found that an arbitrarily high CME is achievable with following conditions: (1) the collision occurs at the horizon(s) of an extremal and non-extremal KNTN black hole, (2) one of the colliding particles has critical angular momentum, and (3) the spin parameter $a\neq0$. We discovered the upper and lower bounds of the Carter constant $K$ for a marginally bound particle with the critical angular momentum in an extremal and non-extremal KNTN black hole. In the equatorial plane, we discovered that there exists intervals for the spin parameter $a$, NUT charge $n$ and specific angular momentum $L$ correspond to the electric charge $Q$ for which not only two marginally bound particles reach the horizon of the extremal KNTN black hole but also the collision of these particles happens at the horizon.
\subsection*{Acknowledgment}
M. Jamil and A. Zakria would like to thank the Higher
Education Commission (HEC), Islamabad for providing financial support
under the project grant no. 20-2166. M. Jamil would also thank the kind hospitality of the Yukawa Institute for Theoretical Physics (YITP), Kyoto University, Kyoto, Japan where this work was initiated. We would also thank the referee for giving useful comments on this work.
\subsection*{Appendix}
The curvature invariants for KNTN metric are given by
\begin{eqnarray}
\mathcal{I}_{1}&=&g^{\mu\nu}R_{\mu\nu}=0,\label{67}\\
\mathcal{I}_{2}&=&R_{\mu\nu}R^{\mu\nu}=\frac{64Q^{4}}{\big(r^{2}+(n+a\cos\theta)^{2}\big)^{6}},\label{68}\\
\mathcal{I}_{3}&=&R_{\mu\nu\rho\sigma}R^{\mu\nu\rho\sigma}=\frac{8(\varepsilon_{1}+\varepsilon_{2}\cos\theta+\varepsilon_{3}\cos^{2}\theta
+\varepsilon_{4}\cos^{3}\theta+\varepsilon_{5}\cos^{4}\theta+\varepsilon_{6}\cos^{5}\theta+\varepsilon_{7}\cos^{6}\theta)}
{\big(r^{2}+(n+a\cos\theta)^{2}\big)^{6}},\label{69}
\end{eqnarray}
where
\begin{eqnarray}
\varepsilon_{1}&=&6n^{8}-12n^{6}Q^{2}+7n^{4}Q^{4}-90n^{6}r^{2}+120n^{4}Q^{2}r^{2}-34n^{2}Q^{4}r^{2}+90n^{4}r^{4}-60
n^{2}Q^{2}r^{4}+7Q^{4}r^{4}\notag\\&&-6n^{2}r^{6}-6M^{2}(n^{6}-15n^{4}r^{2}+15n^{2}r^{4}-r^{6})+12Mr(6n^{6}-Q^{2}r^{4}-5n^{4}
(Q^{2}+4r^{2})\notag\\&&+2n^{2}(5Q^{2}r^{2}+3r^{4})), \label{70}\\ \varepsilon_{2}&=&4an(9n^{6}-17Q^{4}r^{2}-15Q^{2}r^{4}-15n^{4}(Q^{2}+6r^{2})-9M^{2}
(n^{4}-10n^{2}r^{2}+5r^{4})+n^{2}(7Q^{4}\notag\\&&+90Q^{2}r^{2}+45r^{4})+6Mr(15n^{4}+10Q^{2}r^{2}+3r^{4}-10n^{2}(Q^{2}+3r^{2}
))), \label{71}\\ \varepsilon_{3}&=&-2a^2(-45n^{6}+17Q^{4}r^{2}+30n^{4}(2Q^{2}+9r^{2})+45M^{2}(n^{4}-6n^{2}r^{2}+r^{4})-3n^{2}(7Q^{4}+60Q^{2}r^{2}
\notag\\&&+15r^{4})-60Mr(6n^{4}+Q^{2}r^{2}-3n^{2}(Q^{2}+2r^{2}))), \label{72}\\
\varepsilon_{4}&=&4a^{3}n(30n^{4}+7Q^{4}+900r^{2}-30M^{2}(n^{2}-3r^{2})-60Mr(-3n^{2}+Q^{2}+r^{2})-30n^{2}(Q^{2}
\notag\\&&+3r^{2})),\label{73}\\
\varepsilon_{5}&=&-a^4(-90n^{4}-7Q^{4}-60M(6n^{2}-Q^{2})r+90M^{2}(n^{2}-r^{2})+30n^{2}(2Q^{2}+3r^{2})),\label{74}\\ \varepsilon_{6}&=&-12a^{5}n(3M^{2}-3n^{2}+Q^{2}-6Mr), \label{75}\\
\varepsilon_{7}&=&-6a^{6}(M^{2}-n^{2}). \label{76}
\end{eqnarray}
The non-trivial curvature invariants are finite at $\Delta=0$ and infinite at $\Sigma=0$, hence these are coordinate and curvature singularities, respectively.

\end{document}